\documentclass[12pt]{iopart}
\usepackage{graphicx}

\begin{document}

\title[Orbital dynamics in 5 dimensions]{The effect of a fifth large-scale space-time dimension on orbital dynamics}

\author{M B Gerrard and T J Sumner}

\address{Department of Physics, Imperial College London, Prince Consort Road,
London SW7 2BW, UK}
\eads{\mailto{Michael.Gerrard@partnershipsuk.org.uk},
\mailto{t.sumner@imperial.ac.uk}}
\begin{abstract}
A model based on simple assumptions about 4-dimensional space-time
being closed and isotropic, and embedded in a $5^{th}$ large-scale
dimension $r$ representing the radius of curvature of space-time,
has been used in an application of Newton's Second Law to describe a
system with angular momentum. It has been found that the equations
of MOND used to explain the rotation curves of galaxies appear as a
limit within this derivation and that there is a universal
acceleration parameter, $a_o$, with a value, again consistent with
that used by MOND.  This approach does not require {\it
modification} of Newtonian dynamics, only its extension into a fifth
large-scale dimension. The transition from the classical Newtonian
dynamics to the MOND regime emerges naturally and without the
introduction of arbitrary fitting functions, if this 5-dimensional
model is adopted.  The paper also includes the derivation of an
effect in 5-dimensional orbital dynamics which is in reasonable
agreement with the observed Pioneer Anomaly.
\end{abstract}

\maketitle

\section{Introduction}
A discrepancy exists between the extent of observed visible matter
in the universe and that inferred from the motion of galaxies and
clusters of galaxies if Newton's Second Law and Law of Gravitation
are applied. The existence of abundant unobserved (or ``dark")
matter has been proposed by many cosmologists to account for the
discrepancy. An alternative but no less radical explanation is that
either: Newton's Second Law should be modified for the low
accelerations evident in observations of galactic motion; or that
the equations of gravity should be modified for these low
accelerations [1,2]. MOND was first put forward in 1983 [1] and
during the last twenty three years has continued to be consistent
with the growing body of observational data on the rotation curves
of galaxies [3], and has survived a number of studies looking for
consistency with various other effects otherwise requiring dark
matter, such as cosmological models derived from combining a
multitude of observational data [4], to tidal streams from a
companion galaxy to the Milky Way [5].  However, MOND Theory is
empirical in nature, being derived exclusively from correlations
based upon observations and, to date, has eluded derivation from
first principles, although the recent relativistic more fundamental
formulation from Bekenstein [5] offers both MOND and Newtonian
limits.

In this paper we show that the introduction of a $5^{th}$
large-scale dimension to the description of space-time can account
for all aspects of the empirical equations of MOND, provide a
natural transition function $\mu$ between the MOND and Newtonian
regimes, and explain the link between the MOND parameter $a_o$ and
key universal parameters, such as Hubble's Constant. An essential
feature of this approach is that it does {\bf not} require the
principles of Newton's Second Law to be modified, only that they
should be applied in five dimensions of space-time rather than four.

The paper is divided into four sections as follows: formulation of
Newton's Laws in a 5-dimensional space-time; the origin and role of
the acceleration, $a_r$, originating from the presence of a $5^{th}$
dimension; connection to the equations of MOND; and a derivation and
calculation of the Pioneer Anomaly.  The paper concludes with a
short discussion section.

\section{Formulate Newton's Laws in a
5-dimensional space-time} The $5^{th}$ dimension represented by the
parameter $r$ is assumed orthogonal to the three space dimensions
$s(x,y,z)$ and to the time dimension $t$. It is also assumed that
bodies can only propagate in $x, y$, and $z$ and have no access to
the $r$ dimension.  However forces originating from the $r$
dimension can act on bodies in $x, y$, and $z$.  This is illustrated
in figure~1, where only a single space dimension $x$ is shown
together with the $5^{th}$ dimension $r$ forming an $x-r$ plane.

A body at point $o$ is subject to three forces (accelerations): (i)
gravitational acceleration $g_x$ towards a gravitational source mass
$M$ in the direction of negative $x$; (ii) an applied inertial
acceleration $a_x$ in the direction of $x$ positive; and (iii) an
assumed (MOND-like) acceleration $a_r$ in the positive $r$ direction
and orthogonal to $x$. In a world of three space dimensions and no
$5^{th}$ dimension, balance would simply be defined in terms of
component balances in each of the three space dimensions $s(x,y,z)$.
With a $5^{th}$ dimension it is useful to introduce the concept of
branes and look for a condition in which the components in two
mutually orthogonal branes (denoted by symbols $\sigma$ and
$\sigma_\perp$) can be balanced. In figure~1 the angle at which the
branes intersect with the reference axes ($x, r$) is denoted by
$\phi$.

\begin{figure}
\begin{center}
\includegraphics[height=2.5in, clip=]{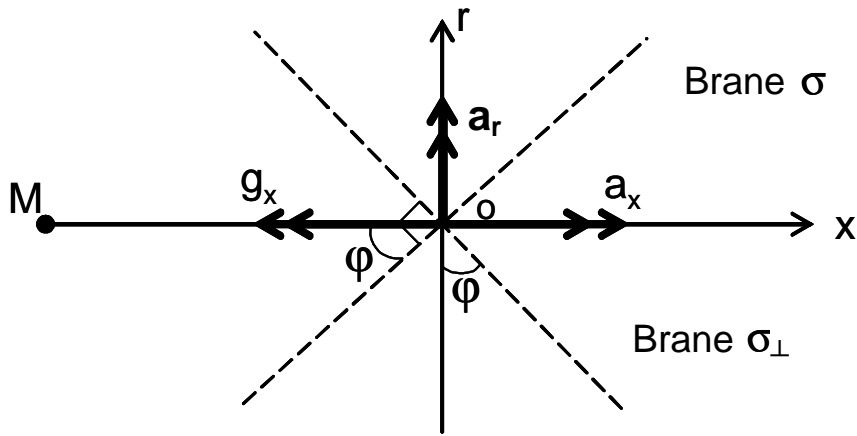}
\caption{Force balance on a body at point $o$ in the $x-r$ plane.}
\end{center}
\end{figure}

Force balance occurs when
\begin{eqnarray}
\mbox{    in brane $\sigma$: } -g_x\cos{\phi}=a_r\sin{\phi}+a_x \cos{\phi} \\
\mbox{    in brane $\sigma_\perp$: }
-g_x\sin{\phi}+a_r\cos{\phi}=a_x\sin{\phi}
\end{eqnarray}

These two equations are simultaneously satisfied if
\begin{equation}
g_x = - a_x \pm ia_r
\end{equation}

Ordinarily it would be assumed that the imaginary part could not be
realised by measurement in `real' space.  However, as will be shown
later this imaginary part can give rise to measurable effects under
certain circumstances.

Before leaving this section we return to the consider the angle
$\phi$, which defines where the brane walls are found.  This angle
is given by
\begin{equation}
\tan{\phi}  = \frac{dr}{dx}
\end{equation}
This angle will not be constant throughout space but will be
determined by the local value of this gradient.

\section{The origin and role of
acceleration, $a_r$, originating from the presence of a $5^{th}$
dimension}
\subsection{The origin of $a_r$}
In a closed universe, which is isotropic on a large-scale, there is
a finite radius of curvature to space-time even remote from
gravitating bodies.  We attribute this underlying parameter, $r_u$,
to our assumed $5^{th}$ dimension. Propagating electro-magnetic
radiation effectively follows a world-line of radius of curvature
$r_u$ and this can be interpreted as a universal acceleration
applying throughout space (see section~4.1) with a value
\begin{equation}
a_r=\frac {c^2}{r_u}
\end{equation}
acting in the direction of positive $r$.\footnote{This acceleration
has, of course, already been empirically associated with the MOND
acceleration parameter $a_o$, through the equation $a_o = c^2/R_c$,
where $R_c$ is the curvature [7]}

The validity of this relationship implies that a closed
4-dimensional space-time is considered to be embedded within a
locally flat 5-dimensional universe.  This is illustrated in
figure~2 from which we now derive $a_r$.

\begin{figure}
\begin{center}
\includegraphics[height=2.5in,clip=]{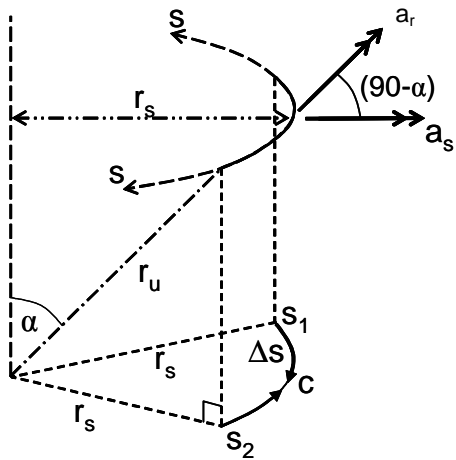}
\caption{Relationship between curvature of space-time $r_u$ and the
acceleration $a_r$.} \label{fig2}
\end{center}
\end{figure}

In figure~2 the space dimension $s$ represents any one of the three
space dimensions $s(x,y,z)$. A single space dimension segment
$s_1s_2$ represents a distance $\Delta s$. $c$ is the speed of light
in vacuo, and the radius of curvature of a (geodesic) path of
electromagnetic radiation in space-time remote from gravitational
fields is $r_u$.  The radius of curvature of single space dimension
segment, $\Delta s$, (remote from gravitational fields) is then
given by $r_s = r_u \sin{\alpha}$. Assuming Newton's second law
applies in 5 dimensions, the centrifugal acceleration acting on
electromagnetic radiation travelling along space dimension segment
$\Delta s$ is given by: $a_s = c^2/r_s$ The value of $a_r$ is now
given by the component of the centrifugal acceleration acting in the
$r$ direction - i.e. in the direction of expansion of the Universe.
Hence:
\begin{equation}
a_r = \frac{c^2}{r_s}\sin{\alpha} = \frac{c^2}{r_u}
\end{equation}

\subsection{Considering the role of $a_r$}
At any point $o$ in space dimension $x$ adjacent to a gravitating
mass $M$, but otherwise empty, there exist two accelerations; (i)
the acceleration due to the gravitational field of $M$, and (ii)
$a_r$ acting in direction of $r$ increasing. This is shown in
figure~3. At the point $o$, a brane can be defined in which there
exists balance between these two accelerations, $g_x$, and $a_r$.
This brane $\sigma$ satisfies the equation:
\begin{equation}
g_x = - a_r \tan{\phi}
\end{equation}
where $\tan{\phi}$ is the local gradient, $dr/dx$, of the brane at
point o relative to the space dimension x.

\begin{figure}
\begin{center}
\includegraphics[height=2.5in, clip=]{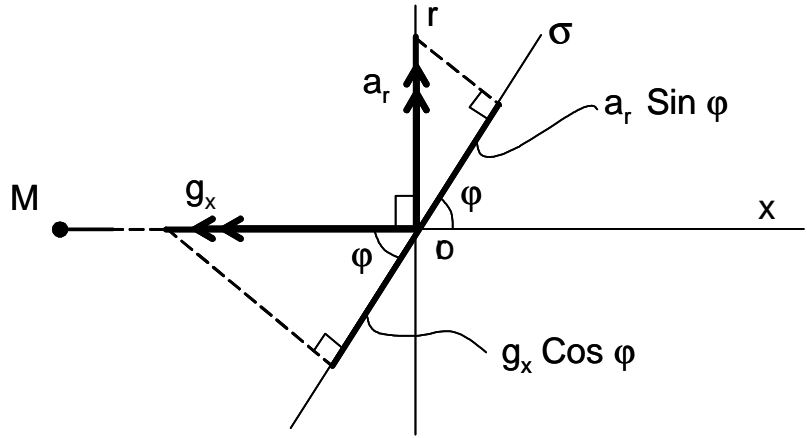}
\caption{Brane ($\sigma$) defined by balance between gravitational
($g_x$) and curvature accelerations ($a_r$).} \label{fig3}
\end{center}
\end{figure}

The brane (defined as $\sigma$ in the diagram) is the locus of all
points $o$ at which there is balance between $g_x$ and $a_r$, i.e.
between accelerations of contraction of the Universe on the one hand
and expansion of the Universe on the other. It can be convenient to
think of this balance condition as a form of ``surface tension"
somewhat akin to the ``rubber-sheet" analogy of General Relativity.
This balance condition is a function of distance $x$, away from the
mass, and this allows a local functional form for $\tan{\phi}(x)$ in
the presence of a gravitating body as:
\begin{eqnarray}
\tan{\phi}=\frac{GM}{a_rx^2}=\frac{r_u}{c^2}\frac{GM}{x^2}=\frac{\partial{r}}{\partial{x}}
\\
\rightarrow r\left( x \right) = r_u \left[ 1 -
\frac{GM}{c^2x}\right]
\end{eqnarray}

An interesting characteristic of the brane $\sigma$ can be noted at
this point simply from the equation of balance between $g_x$ and
$a_r$: the closer is point $o$ to the gravitating mass $M$, the
greater is the gradient $dr/dx$ and the more closely aligned is the
brane $\sigma$ to the $r$-dimension, and as we move further away and
out towards infinity the more the acceleration lying in the plane of
the brane $\sigma$ itself converges to $g_x$.

\section{Connection to the equations of MOND}
The original motivation for MOND was to explain the quasi-flat
rotation curves of galaxies.  Hence to establish if there is a
connection between this work and MOND we need to see how the force
balance in 5 dimensions affects orbiting systems. For an object in a
stable circular orbit (i.e. constant velocity, $v$) in the $x-y$
plane (see figure~4) the centrifugal acceleration in standard
4-dimensional space-time is
\begin{equation}
a_x={\underline \Omega_z}\,\wedge\,{\underline v}
\end{equation}
Adapting this for 5 dimensions requires the introduction of a new
angle, $\theta$, between the angular momentum vector,
$\underline{\Omega}_{zr}$, which now lies in the $z-r$ plane and the
$r$ axis as shown in figure~5. Equation~10 then becomes
\begin{equation}
a_x={\underline \Omega_{zr}}\,\wedge\,{\underline v} =
|\Omega_{zr}|\left( \sin{\theta}+i\cos{\theta} \right)v
\end{equation}
where the imaginary component is in the $r$ direction. \\

The angle $\theta$ will again be locally determined and by contrast
with the variation in the angle $\phi$ which arises from the
dependence of potential energy on position, we associate $\theta$
with the kinetic energy. Replacing potential energy with the
non-relativistic kinetic energy in equation~9 gives\footnote {It is
interesting to note that equation~12 suggests we can also write
$r\left( v \right) = r_u \sqrt{1 - v^2/c^2}$}
\begin{equation}
r\left( v \right) = r_u \left[ 1 - 0.5\frac{v^2}{c^2}\right]
\end{equation}
\begin{figure}
\begin{center}
\includegraphics[height=1.5in]{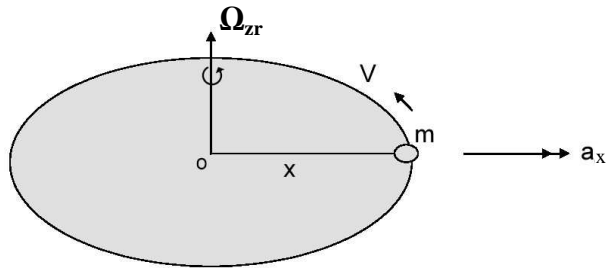}
\caption{Rate of change in direction of velocity.} \label{fig4}
\end{center}
\end{figure}
%\begin{figure}
%\begin{center}
%\includegraphics[height=1.3in]{fig5.eps}
%\caption{Angular velocity in plane $xy$ defined by either
%$\Omega_z$ or $\Omega_r$.} \label{fig5}
%\end{center}
%\end{figure}
\begin{figure}
\begin{center}
\includegraphics[height=1.3in]{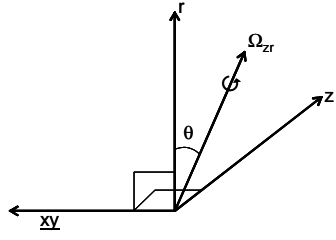}
\caption{Angular velocity vector ${\underline \Omega_{zr}}$ lying in
the plane $zr$.} \label{fig6}
\end{center}
\end{figure}
$|\Omega_{zr}|$ is independent of $\theta$ and can be determined
from the familiar case of ${\underline \Omega_{zr}}$ being aligned
strictly in the direction of $z$ (i.e. $\theta = \pi/2$) when
$|\Omega_{zr}| = v/\rm{\mbox{(radius of orbit)}}$. As for $\theta$
we attribute this to the gradient, $dr/dx$, induced by the inertial
accelerations acting on a body, and this must be calculated from
equation~12 using
\begin{equation}
\frac{dr}{dx}=\frac{dr}{dv}\frac{dv}{dx}
\end{equation}
%\begin{figure}
%\begin{center}
%\includegraphics[height=1.3in]{fig7.eps}
%\caption{The plane of rotation of $m$ lies in $xyr$ space at an
%angle $\theta$ to the plane $xy$.} \label{fig7}
%\end{center}
%\end{figure}
For the body in orbit in figure~4 this gives
\begin{equation}
\frac{dr}{dx}=
-r_u\frac{v}{c^2}\frac{dv}{dx}=-\frac{r_u}{c^2}|\Omega_{zr}|^2x
\end{equation}
However, $|\Omega_{zr}|^2x$ is just the expected Newtonian {\em
inertial} acceleration, $a_N$, and $dr/dx=-\tan{\theta}$ (see
figure~6). Hence
\begin{equation}
\tan{\theta} = \frac{a_N}{a_r}
\end{equation}
\begin{figure}
\begin{center}
\includegraphics[height=1.3in]{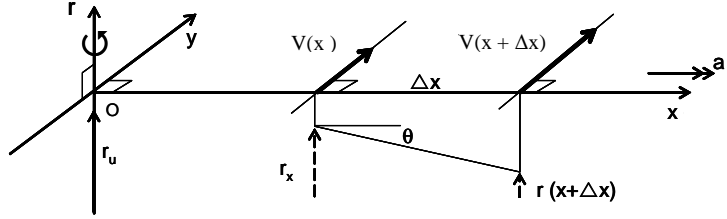}
\caption{: The orbital plane lies in $xyr$ space at an angle
$\theta$ to the $xy$ plane.} \label{fig8}
\end{center}
\end{figure}
An implication of the function $r = r(v)$ is that a full description
of the motion of bodies in three dimensional space $s(x,y,z)$
necessarily requires consideration of a 5-dimensional frame of
reference (i.e. $s(x,y,z)$, $r$ and $t$). The implications of the
equation $\tan{\theta} = a_N/a_r$ are twofold: first, that all
finite angular momentun vectors have a component in the $r$
dimension, and second that the plane of rotation of a body will lie
solely in the three space dimensions $s(x,y,z)$ only if the
centrifugal acceleration of the body tends to zero. One consequence
of these principles is that the rest vector (defined as $\Omega_o$)
for all zero angular velocities must be aligned in the direction of
the $5^{th}$ dimension in the sense of $r$ positive. This is the
same as saying that the default zero angular velocity component lies
in the $5^{th}$ dimension for all matter and energy.  As the {\it
total} angular velocity, \underline{$\Omega$}, increases the vector
becomes increasingly displaced towards the $z$-direction, i.e. being
always normal to the plane of rotation ($xyr$ space in figure~6),
\underline{$\Omega$} subtends an angle $\theta$ away from the
$r$-direction.

Returning to the `centrifugal' accelerations expressed in a
5-dimensional framework, we can now use the expression for
$\tan{\theta}$ in equation 11 which gives
\begin{equation}
a_x=a_N\left[(a_N/a_r)\left( 1 + a_N^2/a_r^2 \right)^{-0.5} +
i\left(1 + a_N^2/a_r^2\right)^{-0.5}\right]
\end{equation}
which, for small $a_N$ ($a_N << a_r$), becomes:
\begin{equation}
a_x= \frac{a_N^2}{a_r} + ia_N
\end{equation}
The real part of this expression recovers the MOND limit if $a_r$ is
identified directly with the $a_o$ parameter from MOND.

For large $a_N$ ($a_N >> a_r$), it becomes:
\begin{equation}
a_x= a_N + ia_r
\end{equation}
the real part of which recovers the Newtonian limit. Hence the real
terms in these equations reproduce the key MOND equations and do
indeed argue for a clear association between this derivation and
MOND. Moreover this analysis provides a form for the transition
function $\mu$ of MOND directly as equation~16\footnote{This
functional form of $\mu$, derived here from first principles, is
precisely one of the fitting functions noted by Milgrom
[http://www.astro.umd.edu/~ssm/mond/faq.html] to give good agreement
with observational data.}. It can be seen from the above derivation
that this transition function is governed ultimately by equation~12,
which is the Lorentz contraction formula approximation for $v<<c$.
In addition the value of $a_r$ is now fixed by equation~5 and in the
next subsection we show that its value is remarkably close to the
value of $a_o$ used by MOND.

Substitution of equation~18 into equation 3, to satisfy the
condition of a force balance in 5-dimensions, shows that $g_x = a_N$
as expected in the Newtonian limit. However if we do the same with
equation 17 there is a residual imaginary term which in the limit
$a_N \rightarrow 0$ becomes $ia_r$.  This appears to be consistent
with a seemless transition at the fringes of galaxies from fixed
orbits to those increasingly governed by the expansion of the
Universe.

\subsection{The value of $a_r$}
To calculate a value for $a_r$ from equation~5 we adopt a
straight-forward cosmology with an expanding universe in which time
increments, $\Delta t$, and space increments, $\Delta s$ can be
related to $r_u$ as shown in figure~7, which gives
\begin{eqnarray}
\Delta t = \frac{r_u}{c} \Delta \alpha \\
\Delta s = r_u \Delta \phi \sin{\alpha}
\end{eqnarray}
\begin{figure}
\begin{center}
\includegraphics[height=3.0in]{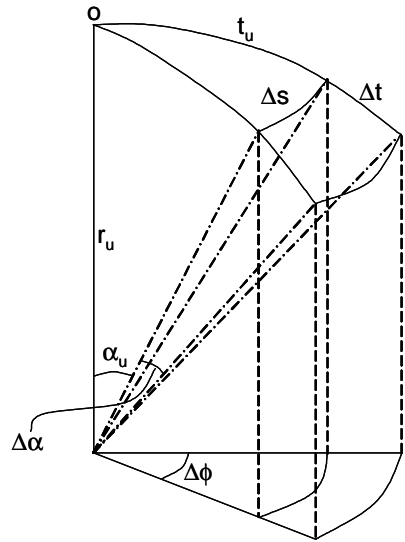}
\caption{: Increments of space ($\Delta s$) and time ($\Delta t$)
are defined by reference to $r_u$.} \label{fig11}
\end{center}
\end{figure}
Consider two galaxies embedded in expanding space-time with zero
peculiar motion; i.e. $\Delta \phi$ is constant.   Their apparent
mutual recession velocity, $v_r$, will be given by $ds/dt$ at
constant $\Delta \phi$ and $r_u$.
\begin{equation}
v_r = \Delta s \frac{c}{r_u} \cot{\alpha}=\Delta s
\frac{c}{r_u}\cot{\frac{ct_u}{r_u}}
\end{equation}
By Hubble's Law this should also be equal to $H\Delta s$ which
provides the link between $H$ and $r_u$
\begin{equation}
H = \frac{c}{r_u} \cot{\frac{ct_u}{r_u}}
\end{equation}
If the value for $H$ is taken to be $71\,\rm{kms}^{-1}\rm{Mpc}^{-1}$
and the age of the universe, $t_u$, is taken to be
$13.7\times10^9\,$years, then equation~22 gives a minimum value for
$r_u = 6.3\times 10^{26}\,$m which substituted into equation~5 gives
a value of $a_r = 1.4 \times 10^{-10}\,$ms$^{-2}$. This is in good
agreement with the MOND parameter $a_o$ derived from observations of
rotation curves of galaxies [1,2], so the connection with MOND is
now clear and in the remaining sections we substitute $a_o$ for
$a_r$\footnote{This means that equations 4 and 7 can be combined and
written as $g_x = - a_o \frac{dr}{dx}$, although there may be good
reasons why a more accurate version is $g_x = - a_o
\frac{dr}{dx}\frac{r_u}{r}$.}.  If the MOND parameter value of
$1.2\times 10^{-10}\,$ms$^{-2}$ is adopted then the implied value
for $r_u$ is $7.5\times 10^{26}\,$m.

\subsection{Galaxy dynamics}
The original motivation for MOND, and hence this work, was in
explaining two aspects of galaxy dynamics; i.e. the flat rotation
curves and the Tulley-Fisher relationship.  Equation~16 allows us to
predict what we expect from the derivation in this paper. Matching
the reals parts of equations 16 and 3 gives
\begin{equation}
\frac{GM\left(x\right)}{x^2}=\frac{v^2}{x}\left[\frac{v^2}{a_ox}\left(1+\frac{v^4}{a_o^2x^2}\right)^{-1/2}\right]
\end{equation}
where $M\left(x\right)$ is the mass enclosed within a radius $x$.
The rotation curve, $v\left(x\right)$, is then given by
\begin{equation}
v=\sqrt{\frac{GM}{\sqrt{2}x}}\left(1+\sqrt{1+\frac{4a_o^2x^4}{\left(GM\right)^2}}\right)^{1/4}
\end{equation}
As an example of the transition region for a system with a
centralised mass distribution, the left-hand panel in figure~8 shows
this function for the solar-system.  The departure from Newtonian
gravity is visually obvious on this plot beyond 300\,AU. The
asymptotic orbital velocity is $\sim\,360$\,m/s.  In the right-hand
panel we show the difference between the Newtonian prediction and
that from equation~24.
\begin{figure}
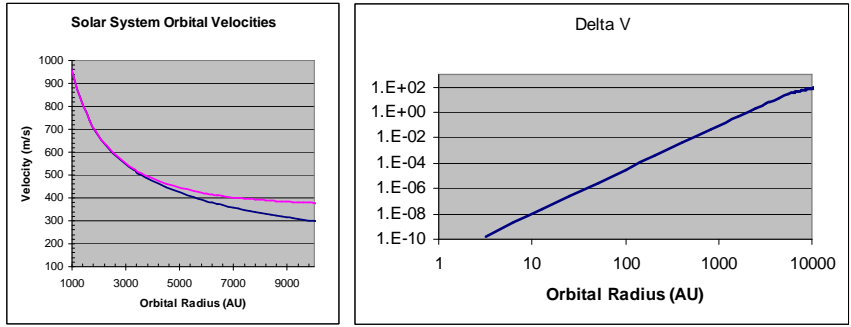

\begin{center}
\includegraphics[height=1.7in]{ssrot.prn}
\includegraphics[height=1.7in]{deltav.prn}
\caption{Transition between Newtonian and MOND regimes for the solar
system.} \label{ss_rot}
\end{center}
\end{figure}
As an example of a galactic rotation curve, figure~9 is plotted for
a distributed mass inferred from measured data from a typical spiral
galaxy.  In figures~8\,\&\,9 the lower curve show the Newtonian
behaviour. It is worth noting that in the current derivation there
are no free parameters from one galaxy to the next, other than
needing to know the mass distribution. Uncertainty in the value of
$a_o$ is a systematic which applies to all galaxies.
\begin{figure}
\begin{center}
\includegraphics[height=1.7in]{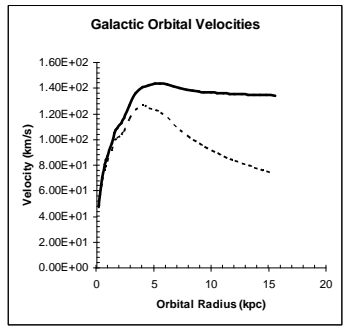}
\caption{An example galaxy rotation curve.} \label{galaxy_rot}
\end{center}
\end{figure}
The asymptotic velocity from equation~24 is
\begin{equation}
v_{asym}^4=a_oGM\left(\infty\right)
\end{equation}
which underlies the Tulley-Fisher relationship.

%\begin{figure}
%\begin{center}
%\includegraphics[height=1.3in]{fig9.eps}
%\caption{Rotation of angular velocity vector $\Omega_{zr}$ with
%increasing acceleration $a_N$.} \label{fig9}
%\end{center}
%\end{figure}

\subsection{Linear Motion}
Equation~16 has been derived for a system with angular momentum and
predicts an observable departure from a Newtonian system. What if we
now consider a system without angular momentum? Combining
equations~9 and 12 we can write
\begin{equation}
\frac{dr}{dx}=\frac{r_u}{mc^2}\left( \frac{dU}{dx}-\frac{dT}{dx}
\right)
\end{equation}
where $U$ is potential energy and $T$ is kinetic energy of a body
with mass $m$.  For a body undergoing purely gravitational
acceleration towards a much larger mass $M$ without any modified
response from the Newtonian expectation this becomes
\begin{equation}
\frac{dr}{dx}=\frac{r_u}{c^2}\left(\frac{GM}{x^2}-a_x
\right)=\frac{1}{a_o}\left(-g_x-a_x\right)=\tan{\phi}
\end{equation}
which is none other than equation 1.  This implies that there are no
components of linear momentum vectors lying in the $r$ dimension.
Moreover if we also impose an energy balance with respect to time,
\begin{equation}
\left(\frac{dU}{dt}+\frac{dT}{dt}\right) = 0
\end{equation}
into which we substitute equations 9 and 12 we get the result
\begin{equation}
g_x = a_N
\end{equation}

Hence there is a balance condition without reference to the
additional dimension $r$ (or $a_o$) and so without departure from
Newtonian principles applied in 4-dimensional space-time. This
implies that MOND-type behaviour is {\it not} to be expected from a
system without angular momentum.

\section{Pioneer Anomaly}
Finally, we consider whether the 5-dimensional orbital dynamics
described in this paper can be used to explain the Pioneer Anomaly
[10].  The acceleration parameter, $a_r$, introduced in section 3.1,
itself has two components: an angular velocity $\Omega_r$ (with
magnitude $c/r$); and $c$ for propagating radiation. The angular
velocity, $\Omega_r$, is orthogonal to $a_r$ and, being a function
of $r$, will vary as $r$ varies.  In particular, $\Omega_r$ will
vary with time for a body moving away from a gravitating mass, $M$,
in accordance with equation 9.  However, to find the complete time
derivative of $\Omega_r$ account must also be taken of the time
dependency of the radius of curvature of space in an expanding
universe as described in sections 3.1 and 4.1.  Hence we start with
the angular velocity $\Omega_s$ that is associated with the
acceleration parameter, $a_s$, for a space dimension interval
$\Delta s$ (see figures 2 and 10).

\begin{equation}
\Omega_s = \frac{c}{r\sin{\alpha}}
\end{equation}

\begin{figure}
\begin{center}
\includegraphics[height=2.5in,clip=]{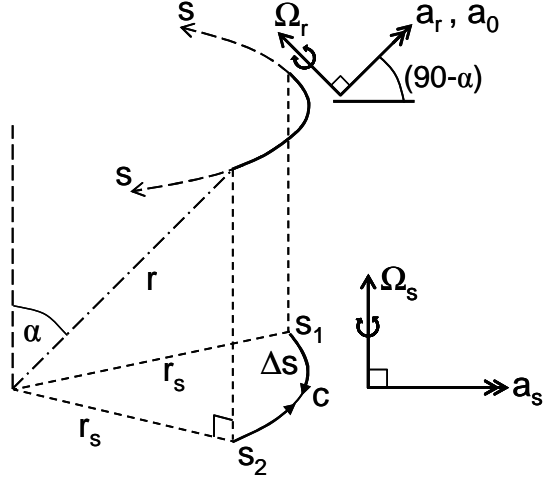}
\caption{Angular velocities in 5-dimensions} \label{fig10}
\end{center}
\end{figure}

Substituting for $r$ from equation 9 (in terms of distance $s$
rather than $x$) and for $\alpha$ from equation 19, the time
derivation of $\Omega_s$ can be expressed as

\begin{equation}
\dot{\Omega}_s =
-\frac{c}{r\sin{\alpha}}\left(\left[\frac{r_u}{r}\right]\left[\frac{g_s}{c^2}\right]\frac{ds}{dt}+\cot{\alpha}\frac{d\alpha}{dt}\right)
\end{equation}
where $g_s$ is the gravitational acceleration due to a mass $M$,
$\frac{ds}{dt}$ is the radial velocity of a body with respect to
$M$, (defined as $v_{radial}$); $\frac{d\alpha}{dt}$ is $c/r$ in the
general case of $r(s)$ rather than $r_u$ in equation 19; $r_u/r\sim
1$; and, from equation 22, $H=(c/r)\cot{\alpha}$, again in the
general case of $r(s)$ rather than $r_u$.  The angular acceleration
$\dot{\Omega}_r$ is the component of $\dot{\Omega}_s$ resolved in
the direction orthogonal to $a_r$. Hence equation 31 can be
approximated by

\begin{equation}
\dot{\Omega_r} = -\frac{1}{r}\left(g_s\frac{v_{radial}}{c} +
Hc\right)
\end{equation}
where $\dot{\Omega_r}$ is the angular acceleration derived in the
first part ($g_sv_{radial}/c$) from motion in the direction of $s$
radially away from the $M$ and, in the second part ($Hc$), from the
consequences of unbound motion (i.e. not within a fixed orbit) that
is subject to the influence of the background expansion of the
universe.  The angular acceleration $\dot{\Omega}_r$ is orthogonal
to $g_s$ and gives rise to a tangential acceleration (i.e an
observed linear acceleration) acting in the direction of $s$
orthogonal to both $\Omega_r$ and $r$ from the relationship

\begin{equation}
\rm{Tangential}\,\rm{acceleration} = r\,\wedge\,\dot{\Omega_r}
\end{equation}

This tangential acceleration appears to be Coreolis-like in nature
within the context of 5-dimensional orbital dynamics - i.e. it
derives from angular accelerations that arise if there is radial
motion ($d(r\sin{\alpha})/dt$) orthogonal to an angular velocity
($\Omega_s$). It is this Coreolis-like acceleration which is
identified with the Pioneer Anomaly and is given the symbol
$a_{PA}$. From equations 32 and 33, $a_{PA}$ can be written:

\begin{equation}
a_{PA} = - g_s\frac{v_{radial}}{c} - Hc
\end{equation}

Equation 34 can be used to predict values for the Pioneer Anomaly
acceleration ($a_{PA}$).  In particular, for Pioneer 11 between
distances of 22\,AU and 32\,AU from the sun; and, for Pioneer 10,
between 40\,AU and 70\,AU [11]. The respective average radial
velocities of the spacecraft at these distances were
$11.8\,$kms$^{-1}$ and $12.6\,$kms$^{-1}$. The figure for the mass
of the sun is taken to be $2\times10^{30}$\,kg and $H =
71$\,kms$^{-1}$Mpc$^{-1}$. The predicted values for $a_{PA}$ are
shown in table~1.

\begin{table}[t]
\caption{Predicted values for the Pioneer Anomaly
acceleration.\label{tab:exp}} \vspace{0.4cm}
\begin{center}
\begin{tabular}{|c|c|}
\hline
{\bf Distance} & ${\bf a_{PA}\times 10^{10}}$ \\
{\bf (AU)} & {\bf (ms$^{-2}$)} \\ \hline \underline {Pioneer 11} &
\\ \hline 22 & 11.7 \\ \hline 27 &  10.1 \\ \hline 32 & 9.2 \\ \hline
\underline {Pioneer 10} &  \\
\hline 40 & 8.5 \\ \hline 50 & 7.9
\\ \hline 60 & 7.6 \\ \hline 70 & 7.4 \\ \hline
{\bf Average} & {\bf 9.0} \\ \hline
\end{tabular}
\end{center}
\end{table}

A number of things can be noted about these results and equation 34:

\begin {itemize}
\item All the figures for $a_{PA}$ lie within the tolerance limits reported
for the Pioneer Anomaly: $8.74\pm1.33\times10^{-10}$\,ms$^{-2}$,
except for those predicted for Pioneer 11 at distances less than
27AU.
\item At the distances considered, the second term ($Hc$) makes
the larger contribution to $a_{PA}$ (in fact, $6.9\times10^{-10}$\,
ms$^{-2}$) and, in this form of the equation, the term does not vary
with distance from the sun, nor with radial velocity ($v_{radial}$).
However, this term only arises after the Pioneer spacecraft have
entered hyperbolic orbits (following their final fly-bys) and their
motion is subject to the Coreolis-like effect described above.
\item If the equation is integrated as
a single continuous function over the distance range 20\,AU to
70\,AU at an average radial velocity of $\sim12.5\,$kms$^{-1}$ and
assuming a hyperbolic orbit throughout, the value for $a_{PA}$
generated is $\sim8.7\times10^{-10}$\,ms$^{-2}$.
\item At
distances of $\sim15$\,AU and with the spacecraft travelling at a
radial velocity $\sim10$\,kms$^{-1}$, but without having entered
hyperbolic orbit, the value of $a_{PA}$ predicted by the equation
remains $\sim9\times10^{-10}$\,ms$^{-2}$ as the loss of the second
term which falls away for a bounded orbit, is made up for by the
stronger gravitational acceleration of the sun in the first term.
\item As radial velocities decline to zero in bound orbits, so will $a_{PA}$.
\item The equation only takes into account a single space
dimension ($s$) - the radial distance from $M$. This prevents it
from describing the full 3-dimensional motion of the Pioneer
spacecraft as they travel through and out of the solar system. In
addition, the extent to which the transition from bound to unbound
orbits progressively introduces the second term is not addressed.  A
full 5-dimensional derivation of the equation should correct this
short-coming, but is beyond the scope of this paper.
\item In
circumstances where the hyperbolic orbit term does apply, the scale
link between $a_{PA}$ and the MOND parameter $a_0$, is apparent from
the relationship $Hc = a_0\cot\alpha$ (derived from equations 5 and
22) where $\cot\alpha = 5.7$ for the current age of the universe.
\end{itemize}
The level of agreement between the values for $a_{PA}$ predicted by
equation 34 and those reported seems to be reasonably good, albeit
that the reported value is interpreted to be constant whereas
equation 34 predicts a range of possible values that depend on
several key parameters, such as the radial velocity of the Pioneer
spacecrafts.

\section{Discussion}

A model based on simple assumptions about 4-dimensional space-time
being closed and isotropic, and embedded in a $5^{th}$ large-scale
dimension $r$ representing the radius of curvature of space-time,
has been used to derive the balance conditions for a system with
angular momentum.  It has been found that the equations of MOND
explaining the rotation curves of galaxies appear as a limit within
this derivation and that there is a universal acceleration
parameter, $a_o$, with a well defined value, again consistent with
that used by MOND. In addition this derivation provides, for the
first time, a derived transition formula between the Newtonian and
MOND limits. The following positive points should be noted:-

\begin{itemize}
\item Derivation of the Tulley-Fisher relationship from first
principles.
\item Derivation of the MOND acceleration parameter
($a_o$) from key cosmological constants.
\item Identifying $a_o$ as a
universal acceleration of expansion acting everywhere in opposition
to gravity (as the universal acceleration of contraction), so that
the two opposite accelerations jointly define conditions of
``tension'' within branes occupied by matter and energy.
\item Using, as foundation stones, key relationships
from General and Special Relativity, in particular:
\item Identifying the Lorentz contraction formula as governing the
smooth transition from MOND to classical Newtonian regimes, as well
as from classical Newtonian to relativistic regimes - so avoiding
the need for an arbitrary MOND transition function ($\mu$).
\item Generating equations of motion which appear to describe a natural
transition from fixed orbits to those, on the fringes of galaxies
and clusters of galaxies, which are affected by the expansion of the
universe.
\item A possible derivation of the Pioneer Anomaly using
5-dimensional orbital dynamics including a calculation involving key
cosmological constants with no free parameters.
\end{itemize}

The principal objection to introducing a fifth large-scale dimension
into models of the universe is generally taken to be a lack of
direct observational evidence to support its existence. However, the
approach adopted in this paper uses a similar premise to many other
papers in assuming the fifth dimension is quite unlike the three
space (i.e. visible) dimensions $s(x,y,z)$, in that: first, it has
to be treated as mathematically imaginary; and second, it does not
represent a degree of freedom of motion for matter or energy.  In
consequence, the case for a fifth large-scale dimension must be
inferred rather than directly observed. Two potential derivations of
MOND actually proposed by Milgrom - one based on modified inertia
and one based on branes in higher dimensional space-time - are, to
an extent, both reflected in the 5-dimensional model proposed in
this paper. However, a key conclusion of this paper, which is at
odds with these two derivations is that departures from classical
Newtonian dynamics - as observed within four-dimensional space-time
- will only be seen in systems with angular velocity or, as in the
case of the Pioneer Anomaly, where angular accelerations arise
within 5-dimensions.

\end{document}